\documentclass[aps,prx,reprint,superscriptaddress]{revtex4-1}
\usepackage{array,graphicx,amsmath,bm,amsthm,amssymb,setspace,amsfonts}
\usepackage{enumerate}
\usepackage{subfigure}

\begin{document}


\title{Nonequilibrium steady state of biochemical cycle kinetics under non-isothermal conditions}


\author{Xiao Jin}

%\thanks{}

\affiliation{Beijing International Center for Mathematical Research (BICMR), Peking University, Beijing 100871,
PRC}

\affiliation{School of Mathematical Sciences, Peking University, Beijing, 100871, PRC}
\author{Hao Ge}
\email[Electronic address:]{  haoge@pku.edu.cn}
\affiliation{Beijing International Center for Mathematical Research (BICMR), Peking University, Beijing 100871,
PRC}
\affiliation{Biodynamic Optical Imaging Center (BIOPIC), Peking University, Beijing 100871, PRC}


\date{\today}

\begin{abstract}
Nonequilibrium steady state of isothermal biochemical cycle kinetics has been extensively studied, but much less investigated under non-isothermal conditions. However, once the heat exchange between subsystems is rather slow, the isothermal assumption of the whole system meets great challenge, which is indeed the case inside many kinds of living organisms. Here we generalize the nonequilibrium steady-state theory of isothermal biochemical cycle kinetics, in the master-equation models, to the situation in which the temperatures of subsystems can be far from uniform. We first obtain a new thermodynamic relation between the chemical reaction rates and thermodynamic potentials under such a non-isothermal circumstances, which immediately implies simply applying the isothermal transition-state rate formula for each chemical reaction in terms of only the reactants' temperature, is not thermodynamically consistent. Therefore, we mathematically derive several revised reaction-rate formulas which not only obey the new thermodynamic relation but also approximate the exact reaction rate better than the rate formula under isothermal condition. The new thermodynamic relation also predicts that in the transporter system with different temperatures inside and outside the membrane, the net flux of the transported molecules can possibly even go against the temperature gradient in the absence of the chemical driving force.
\end{abstract}
%\keywords{}

\maketitle

\section{Introduction}

Biochemical cycle kinetics under nonequilibrium steady-state condition widely present in cellular activities and are crucial for the utilization of biological functions, such as metabolism, signal transduction, energy transduction and so on \cite{Hill1975,hill2012free,QH2007,Ge2012stochastic,Tu2012,Tu2015,Wo2016}. The thermodynamic analysis of such a mesoscopic system with stochastic dynamics has already been thoroughly investigated under isothermal condition \cite{cox1950statistical,Hill1975,hill2012free,Schnakenberg1976network}, which partly motivated the emergence of stochastic thermodynamics \cite{Ge2010,Seifert2012stochastic,Zhang2012stochastic,Ge2012stochastic}.

The isothermal assumption is only valid when the timescale of the biochemical kinetics is much slower than the timescale of heat exchange within the system. Once the system under investigation is divided into several subsystems, and if the heat exchange between them is sufficiently slow compared to the biochemical kinetics that one is interested in, the temperatures can only be well defined for each subsystem, which can be far from uniform \cite{suzuki2007microscopic,vale1980effect}. For example, in living cells\cite{casas2003temperature,Lukin_nature2013}, the transporter protein across the cell membrane faces both the intracellular and extracellular components of the cell, which are possibly under different temperatures \cite{bedeaux2008measurable}. Hence such a mesoscopic chemical kinetic system is driven not only by the chemical potential differences but also by the temperature gradient across the cell membrane.

Either under the isothermal or non-isothermal circumstances, the mesoscopic biochemical cycle kinetics can be modeled by the master equations. The theory of stochastic thermodynamics in terms of the master-equation model has already been well developed, and the entropy production rate is expressed in terms of the transition rates between discrete states\cite{Seifert2012stochastic,Zhang2012stochastic,Ge2012stochastic}. After taking ensemble average, the mesoscopic stochastic thermodynamics should be consistent with the framework of macroscopic nonequilibrium thermodynamics \cite{Bridgman,Prigogine}. Such a consistency under isothermal condition has already been illustrated \cite{Hill1975,Ge2013}, but not for nonisothermal situations yet.

%The beauty of stochastic thermodynamics is that several central quantities in nonequilibrium thermodynamics can be defined independent of the concrete physical or chemical systems that the stochastic dynamics describes, such as the entropy of the system, the entropy production and the entropy exchange with the environment. In the next step, the physical meanings and significance of these defined thermodynamic quantities should be investigated case by case, in order to get deeper and more concrete understandings of the system.

Here, we carry out a detailed and comprehensive theoretical analysis for nonequilibrium steady state of biochemical cycle kinetics under non-isothermal condition. Starting from a four-state model of the transporter protein across the cell membrane, we obtain a new thermodynamic relation between the reaction rates in the master-equation model and the thermodynamic potentials of discrete chemical states that are involved, based on the consistency between mesoscopic stochastic thermodynamics and macroscopic nonequilibrium thermodynamics. Such a thermodynamic relation is not satisfied if one just simply applies the isothermal transition-state rate formula for each reaction in terms of only the reactants' temperature. Instead we mathematically derive several revised reaction-rate formulas, which not only obey the new thermodynamic relation but also approximate the exact reaction rate better than the isothermal one. The new thermodynamic relation also predicts that the net flux of the transported molecules in the four-state transporter model can even go against the temperature gradient across the cell membrane in the absence of chemical driving force, which does not violate the Second Law of Thermodynamics. Fluctuation theorems are also derived for the four-state model under non-isothermal condition.

We then generalize all the obtained results to a more complicated six-state transporter model, as well as general master-equation models, based on the existing cycle-decomposition theory of the master-equation model \cite{Jiang2004}. We derive the same new thermodynamic relation for each kinetic cycle of the system and validate its consistency with the revised reaction-rate formulas. Various fluctuation theorems are also derived for each kinetic cycle.

%\section{Entropy production rate and a new thermodynamic relation}

\section{A four-state transporter model under nonisothermal condition}

\begin{figure}[hbt]
\centering
\includegraphics[width=7cm]{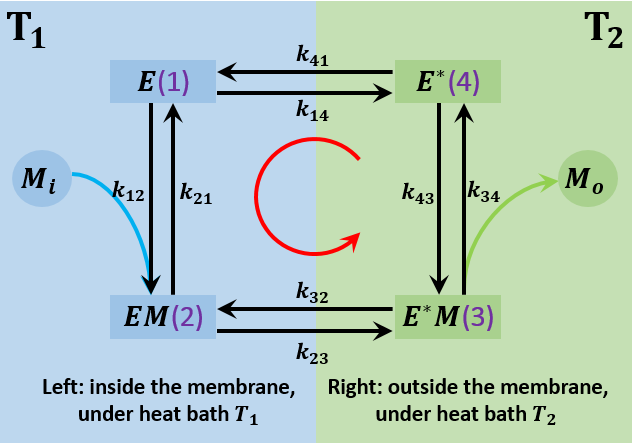}
\caption{Four-state model of the transporter protein across the membrane.To simplify the notations, we denote $E$ as state $1$, $EM$ as state $2$, $E^{*}M$ as state $3$ and $E^*$ as state $4$. $k_{ij}$ is the first-order or pseudo-first-order reaction constants from state $i$ to state $j$. For instance, $k_{12}=k^0_{12}[M_i]$, in which $k^0_{12}$ is the second-order reaction constant and $[M_i]$ is the concentration of $M_i$. Similarly $k_{43}=k^0_{43}[M_o]$.}\label{four-state-model}
\end{figure}

We start from a four-state model of transporter protein across the membrane (Fig. \ref{four-state-model}). The nonequilibrium analysis of this model under isothermal condition can be traced back to T.L.Hill's pioneering work \cite{Hill1975,hill2012free}. $E$ and $E^{*}$ are designated as two conformations of the transporter protein facing the inside and outside of the cell membrane respectively. The intracellular temperature is $T_{1}$ while the extracellular temperature is $T_2$. When a counterclockwise cycle of the reaction diagram in Fig. \ref{four-state-model} is completed, the transporter converts one molecule of $M_{i}$ from the inside to one molecule of $M_{o}$ at the outside.

Thermodynamic equilibrium means both thermal equilibrium $T_1=T_2$ and chemical equilibrium, i.e. the chemical potentials inside and outside the membrane are equal to each other ($\mu_{M_{i}}=\mu_{M_{o}}$). Here the chemical potential should be replaced by the electrochemical potential gradient including the membrane potential effect if the transported molecule is not electrically neutral.

\subsection{Entropy production and a new thermodynamic relation}

In stochastic thermodynamics \cite{seifert2008stochastic,seifert2005entropy}, once a counterclockwise cycle is completed in Fig. \ref{four-state-model}, the entropy production is
\begin{equation}
e_p=k_{B}\ln\left(\frac{k_{12}k_{23}k_{34}k_{41}}{k_{14}k_{43}k_{32}k_{21}}\right).\label{eq1}
\end{equation}

On the other hand, the balance equation of entropy in nonequilibrium thermodynamics tells that $e_p=\Delta S-\Delta S_e$, in which $\Delta S$ is the entropy change of the transporter system in Fig. \ref{four-state-model} after the counterclockwise cycle, and $\Delta S_e$ is the entropy change of the medium due to heat dissipation.

The resulted entropy change of the transporter system during this cycle is just the entropy difference between the molecules $M_o$ and $M_i$, i.e.

\begin{equation}
\Delta S=S_{M_o}(T_2)-S_{M_i}(T_1).
\end{equation}

The entropy change of the medium due to heat dissipation is
$\Delta S_e=-\left(\frac{Q_1}{T_1}+\frac{Q_2}{T_2}\right)$ \cite{Prigogine}, in which $Q_i$ is the heat dissipated to the $i$-th bath with temperature $T_i$, $i=1,2$.

%$dS=d_{e}S+d_{i}S=0$, $d_{i}S=epr=J_{C}k_{B}\ln\gamma$, where cycle flux $J_C$ is the average number of occurrences of the cycle, and

\begin{figure}[hbt]
\centering
\includegraphics[width=7cm,height=5cm]{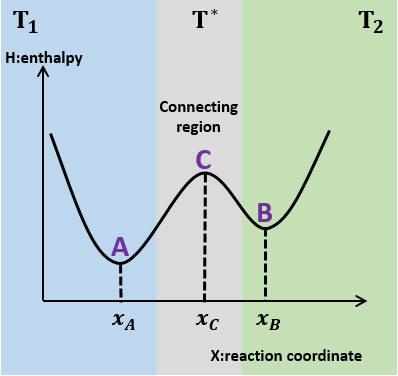}\label{2}
\caption{The enthalpy surface $H(x)$ with two local minima $A$ and $B$. $x$ is the reaction coordinate. Escape from A to B occurs via rate $k_{AB}$. We assume that the temperature of the region belonging to the left-hand side of $X_c$ is $T_1$, while on the right hand side of $X_c$ is $T_2$. The temperature varies in a region around $X_c$, which is called here as the connecting region.}\label{barrier-crossing}
\end{figure}

In order to carry out more detailed analysis of the heat dissipation $Q_1$ and $Q_2$, we consider the barrier-crossing picture for each reaction (Fig. \ref{barrier-crossing}). Transition state $C$ lies at the saddle point of the potential energy surface along the reaction coordinate $x$, while the reactant $A$ and product $B$ are around the two different local minima. Here in Fig. \ref{barrier-crossing} we choose enthalpy as the potential energy, which is typically used for chemical reactions.

During the single reaction $A\rightarrow B$, the heat absorbed from the heat bath at the left-hand side of the transition state with temperature $T_1$ is $H_C-H_A$, while the heat released to the other heat bath at the right-hand side of the transition state with temperature $T_2$ is $H_C-H_B$. Here $H_i$ denotes the averaged enthalpy of the chemical state $i$, $i=A,B,C$.

Therefore, during the completion of a counterclockwise cycle in Fig. \ref{four-state-model}, the heat released to the two heat baths with temperature $T_1$ and $T_2$ are $Q_1=H_{14}^{*}-H_{23}^{*}+H_{M_i}$ and $Q_2=H_{32}^{*}-H_{41}^{*}-H_{M_o}$ respectively, in which $H_{ij}^{*}$ is the averaged enthalpy of the transition state along the reaction coordinate from state $i$ to state $j$. Clearly $H_{ij}^*=H_{ji}^*$.

Thus
\begin{eqnarray}\label{ep_eq}
e_p&=&\Delta S+\frac{Q_1}{T_1}+\frac{Q_2}{T_2}\nonumber\\
&=&\frac{H_{14}^{*}-H_{23}^{*}+\mu_{M_i}}{T_1}+\frac{H_{32}^{*}-H_{41}^{*}-\mu_{M_o}}{T_2}.
\end{eqnarray}

%\subsection{A new thermodynamics relation}

Combined with the definition of entropy production from stochastic thermodynamics [Eq. (\ref{eq1})], we obtain a new relation between reaction rates and thermodynamic potentials of the involved chemical states under non-isothermal condition
\begin{eqnarray}
&&k_{B}\ln\left(\frac{k_{12}k_{23}k_{34}k_{41}}{k_{14}k_{43}k_{32}k_{21}}\right)\nonumber\\
&=&\frac{H_{14}^{*}-H_{23}^{*}+\mu_{M_i}}{T_1}+\frac{H_{32}^{*}-H_{41}^{*}-\mu_{M_o}}{T_2}.
\label{e1}
\end{eqnarray}

If $T_1=T_2$, this relation reduces to the familiar one under isothermal condition \cite{Hill1975,hill2012free}:
\begin{equation}\label{T_1=T_2}
k_{B}\ln\left(\frac{k_{12}k_{23}k_{34}k_{41}}{k_{14}k_{43}k_{32}k_{21}}\right)=\frac{\mu_{M_i}-\mu_{M_o}}{T}.
\end{equation}

In order to keep the nonequilibrium steady state (NESS), there must be an external device responsible for transferring one molecule of $M_o$ from outside back to one molecule of $M_i$ inside, after a counterclockwise cycle is completed \cite{Ge2013}. During this procedure, the classical Clausius inequality tells that
\begin{equation}
S_{M_i}(T_1)-S_{M_o}(T_2)=\Delta S'\geq-\left(\frac{\tilde{Q_1}}{T_1}+\frac{\tilde{Q_2}}{T_2}\right)=\Delta S_e^{'},
\end{equation}
where $\tilde{Q_i}$ is the heat released to the heat bath with temperature $T_i$ during the regenerating process operated by the external device.

Hence the minimal entropy change of the medium due to heat dissipation during the regenerating process becomes
\begin{equation}
\min\{-\Delta S_e^{'}\}=\min\left\{\frac{\tilde{Q_1}}{T_1}+\frac{\tilde{Q_2}}{T_2}\right\}=S_{M_o}(T_2)-S_{M_i}(T_1).
\end{equation}

Finally during the whole process (counterclockwise cycle added with the regenerating process), we have
\begin{eqnarray}
e_p&=&\Delta S+\left(\frac{Q_1}{T_1}+\frac{Q_2}{T_2}\right)\nonumber\\
&=&\left(\frac{Q_1}{T_1}+\frac{Q_2}{T_2}\right)+\min\left\{\frac{\tilde{Q_1}}{T_1}+\frac{\tilde{Q_2}}{T_2}\right\},
\end{eqnarray}
which is exactly the meaning of entropy production at steady state: all dissipated into heat.

\subsection{The transmembrane flux of transported molecules}

%\subsection{Cyclic case}

According to Eq. (\ref{ep_eq}), the entropy production during a counterclockwise cycle in Fig. \ref{four-state-model} can be rewritten as
\begin{eqnarray}
e_p=(H_{14}^{*}+\mu_{M_i}-H_{23}^{*})\left(\frac{1}{T_1}-\frac{1}{T_2}\right)+\frac{1}{T_2}(\mu_{M_i}-\mu_{M_o}).\nonumber
\end{eqnarray}

It is indeed called the thermodynamic driving force for the biochemical cycle \cite{Hill1975,hill2012free}. We can define the chemical driving force as $F_{chem}=\frac{1}{T_2}(\mu_{M_i}-\mu_{M_o})$, and the thermal driving force as $F_{thermo}=(H_{14}^{*}+\mu_{M_i}-H_{23}^{*})\left(\frac{1}{T_1}-\frac{1}{T_2}\right)$. So we have $e_p=F_{chem}+F_{thermo}$. Thermal equilibrium is $T_1=T_2$, which implies
$F_{thermo}=0$, while chemical equilibrium is $\mu_{M_i}=\mu_{M_o}$, which implies $F_{chem}=0$.

There is another way for decomposing $e_p$ into the chemical and thermal driving forces $F_{chem}$ and $F_{thermo}$, i.e. let $F_{chem}=\frac{1}{T_1}(\mu_{M_i}-\mu_{M_o})$ and $F_{thermo}=(H_{14}^{*}+\mu_{M_o}-H_{23}^{*})\left(\frac{1}{T_1}-\frac{1}{T_2}\right)$. Either decomposition will give the same conclusions. So throughout this paper, we keep using the former decomposition.

On the other hand, the positivity of $e_p$ is equivalent to the counterclockwise direction of the net cycle flux in Fig. \ref{four-state-model} \cite{Hill1975,Jiang2004}. Hence, when the chemical force $F_{chem}$ is zero, the direction of the net flux $J_C$, which is the averaged occurrence of the counterclockwise cycle per unit time, is determined only by the thermal driving force $F_{thermo}$, i.e. we have

\begin{equation}\label{againstflux}
J_{C}\cdot F_{chem}=J_{C}\cdot (H_{14}^{*}+\mu_{M_i}-H_{23}^{*})\left(\frac{1}{T_1}-\frac{1}{T_2}\right)>0.
\end{equation}

It indicates that once $(H_{14}^{*}+\mu_{M_i}-H_{23}^{*})>0$, $J_{C}\cdot \left(\frac{1}{T_1}-\frac{1}{T_2}\right)>0$, which implies that the transmembrane flux of molecules goes against the temperature gradient. It seems not meet our intuition but it really does not violate the Second Law, because the thermal driving force $F_{chem}$ here is not $\left(\frac{1}{T_2}-\frac{1}{T_1}\right)$ but should be $(H_{14}^{*}+\mu_{M_i}-H_{23}^{*})\left(\frac{1}{T_1}-\frac{1}{T_2}\right)$.

Such a counterintuitive phenomenon can not occur in a single reaction. Consider the single chemical reaction $A\rightleftharpoons B$ in Fig. \ref{barrier-crossing}, $A$ is at temperature $T_1$, and $B$ is at temperature $T_2$.

The thermodynamic relation in this single chemical reaction satisfies (see Section I of Supplementary Material \cite{supplementary2016material})
\begin{eqnarray}
&&k_B\ln\left(\frac{k_{AB}[A]}{k_{BA}[B]}\right)\nonumber\\
\nonumber
&=&(H_{AB}^{*}-\mu_A)\left(\frac{1}{T_{2}}-\frac{1}{T_{1}}\right)+\frac{1}{T_2}(\mu_{A}-\mu_{B}),
\end{eqnarray}
which has the same sign with the net flux $J=k_{AB}[A]-k_{BA}[B]$.

When the chemical driving force $\frac{1}{T_2}(\mu_{A}-\mu_{B})$ vanishes, we have
\begin{equation}
k_B\ln\left(\frac{k_{AB}[A]}{k_{BA}[B]}\right)=(H_{AB}^{*}-\mu_A)\left(\frac{1}{T_{2}}-\frac{1}{T_{1}}\right),
\end{equation}
followed by
\begin{equation}
J\cdot\left(\frac{1}{T_2}-\frac{1}{T_1}\right)>0,
\end{equation}
since $H_{AB}^{*}-\mu_A$ is always positive. It indicates that the net flux of molecules always follows the temperature gradient in the absence of chemical driving forces.

\subsection{Fluctuation theorems}

It has already been shown that the stochastic net number of occurrences $\nu(t)$ of counterclockwise cycle in Fig. \ref{four-state-model} up to time $t$ satisfies the detailed fluctuation theorem \cite{Ge2012stochastic}

\begin{equation}\label{fluctuations}
\frac{P(\nu(t)=k)}{P(\nu(t)=-k)}=\gamma^k,
\end{equation}
for any integer $k$, in which $\gamma=\left(\frac{k_{12}k_{23}k_{34}k_{41}}{k_{14}k_{43}k_{32}k_{21}}\right)$.

It is followed by \cite{Ge2012stochastic}
$$\langle e^{-\lambda \nu(t)}\rangle=\langle e^{-(\log\gamma-\lambda)\nu(t)}\rangle,$$
for any real $\lambda$.

Therefore, for any physical quantity associated with the counterclockwise cycle, i.e. $W(t)=C\cdot \nu(t)$ for any constant $C$, we can write
\begin{equation}\label{FT_eq1}
\frac{P(W(t)=w)}{P(W(t)=-w)}=\gamma^{w/C},
\end{equation}
for any $w$ in which $w/C$ is an integer, and
\begin{equation}\label{FT_eq2}
\left\langle e^{-\lambda W(t)}\right\rangle=\left\langle e^{-\left(\frac{\log\gamma}{C}-\lambda\right)W(t)}\right\rangle.
\end{equation}

For instance, in the four-state model of transporter in Fig. \ref{four-state-model}, there are at least three thermodynamic quantities that we are interested in: $e_p$, $F_{chem}$ and $F_{thermo}$. Fluctuation theorems Eq.~(\ref{FT_eq1}) and Eq.~(\ref{FT_eq2}) hold for each of them.

\subsection{A more general form of the new thermodynamic relation}

In the high dimensional case, Barezhkovskii et al.\cite{berezhkovskii2005one} have showed that a one-dimensional reaction coordinate exists for any system whose transition rate is described by Langer's multidimensional generalization of the one-dimensional Kramers' theory of diffusive barrier crossing.
Indeed, we can derive a more general form of the new thermodynamic relation [Eq. (\ref{e1})], regarding the temperature as a function of the one-dimensional reaction coordinate without any further assumption or simplification. During the completion of a counterclockwise cycle in Fig. \ref{four-state-model}, the entropy production (see Section II of Supplementary Material for details \cite{supplementary2016material})
\begin{eqnarray}
e_p&=&-\int_{x_1}^{x_2}\frac{H_x'(\chi_{12}(x))}{T(x)}dx-\int_{x_2}^{x_3}\frac{H_x'(\chi_{23}(x))}{T(x)}dx\nonumber\\
&&-\int_{x_3}^{x_4}\frac{H_x'(\chi_{34}(x))}{T(x)}dx-\int_{x_4}^{x_1}\frac{H_x'(\chi_{41}(x))}{T(x)}dx\nonumber\\
&&+\frac{\mu_{M_i}}{T_1}-\frac{\mu_{M_o}}{T_2}\nonumber\\
&=&-\oint\frac{H_x'(x)}{T(x)}dx+\frac{\mu_{M_i}}{T_1}-\frac{\mu_{M_o}}{T_2},
\end{eqnarray}
in which $\chi_{ij}(x)$ is the one-dimensional reaction coordinate along the transition from state $i$ to state $j$, $H(x)$ is the mean potential energy at $x$ and $x_i$ is the coordinate of the state $i$. $\oint$ is just the abbreviation for the integral around a cycle.

Thus
\begin{eqnarray}\label{e2}
&&k_B\ln\left(\frac{k_{12}k_{23}k_{34}k_{41}}{k_{14}k_{43}k_{32}k_{21}}\right)\\
&=&-\oint\frac{H_x'(x)}{T(x)}dx+\frac{\mu_{M_i}}{T_1}-\frac{\mu_{M_o}}{T_2}.\nonumber
\end{eqnarray}

Notice that in the isothermal case with only one heat bath, the first term on the right-hand side of Eq.~(\ref{e2}) vanishes, hence Eq.~(\ref{e2}) reduces to Eq.~(\ref{T_1=T_2}). And once the effect of connecting region on reaction rates is nearly neglectable, Eq.~(\ref{e2}) is reduced to Eq.~(\ref{e1}).

\section{Thermodynamically consistent revised reaction-rate formulas}

\subsection{Simply applying the isothermal transition-state rate formula}

The celebrated transition-rate rate formula is derived under isothermal condition\cite{Hanggi1990,Zhou2010}, and that is
\begin{equation}
k=\kappa\frac{k_{B}T}{h}e^{-\frac{\Delta G^{\ddag}}{k_{B}T}}=\kappa\frac{k_{B}T}{h}e^{-\frac{\Delta H^{\ddag}}{k_{B}T}}e^{\frac{\Delta S^{\ddag}}{k_B}},
\end{equation}
in which $\Delta G^{\ddag}$, $\Delta H^{\ddag}$ and $\Delta S^{\ddag}$ are the free energy, enthalpy and entropy of activation respectively.

Applying the isothermal transition-state rate formula for each reaction in terms of only the reactants' temperature, one can arrive at (see Section III of Supplementary Material \cite{supplementary2016material}):
\begin{eqnarray}
&&k_{B}\ln\left(\frac{k_{12}k_{23}k_{34}k_{41}}{k_{14}k_{43}k_{32}k_{21}}\right)\nonumber\\
&&=\frac{H_{14}^{*}-H_{23}^{*}+\mu_{M_i}}{T_1}+\frac{H_{32}^{*}-H_{41}^{*}-\mu_{M_o}}{T_2}+\Delta.
\label{limitation}
\end{eqnarray}
in which $\Delta=-S_{14}^{\ddag}(T_1)+S_{41}^{\ddag}(T_2)-S_{32}^{\ddag}(T_2)+S_{23}^{\ddag}(T_1)$. $S_{ij}^{\ddag}$ is the entropy of the transition state along the reaction coordinate from state $i$ to state $j$.

There is no theoretical support for the always vanishing of $\Delta$, hence it is contradictive to the new thermodynamic relation [Eq.~(\ref{e1})].

The term $\Delta$ in Eq.~(\ref{limitation}) emerges because the transition states for the forward and backward reactions are attributed different temperatures. It is indeed not reasonable. The transition state should have its own temperature $T^*$, and we need a revised reaction rate formula under the nonisothermal condition.

\subsection{The revised reaction rate formulas}

The reaction rate from reactant $A$ to product $B$, crossing the transition state $C$ in Fig. \ref{barrier-crossing}, can be defined as the reciprocal of the mean first passage time from one local minimum $x_A$ of the enthalpy surface to the other local minimum at $x_B$.

Consider the overdamped Langevin dynamics with the potential energy surface $H(x)$ \cite{Hanggi1990}:
\begin{equation}
d\mathbf{X}_{t}=-\frac{\nabla H(\mathbf{X}_{t})}{\eta}dt+\sqrt{\frac{2k_{B}T(\mathbf{X}_{t})}{\eta}}d\mathbf{B}_{t},
\end{equation}
in which the temperature $T$ is a function of $x$, and the local Einstein relation holds.

\subsubsection{One dimensional case}

\begin{figure*}[hbt]
\centering
\subfigure[]{\includegraphics[width=5.5cm,height=5cm]{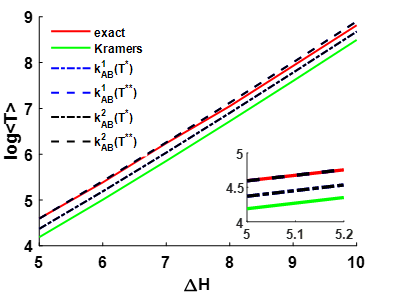}}
\subfigure[]{\includegraphics[width=5.5cm,height=5cm]{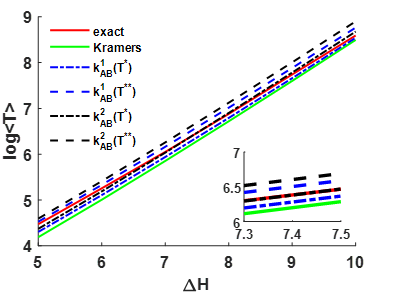}}
\subfigure[]{\includegraphics[width=5.5cm,height=5cm]{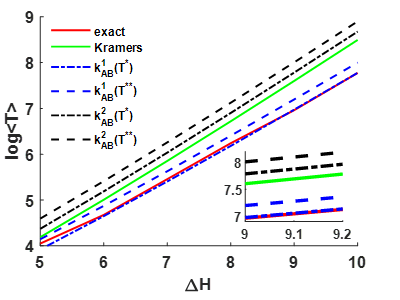}}
\caption{Validation of the revised rate formulas under non-isothermal condition in one dimensional case. The exact reaction rate is obtained via analytical expressions. The enthalpy function used is $H(x)=c*(\frac{1}{2}x^4-x^2)$, $\Delta H=0.5c$. $T^*/T_1=1.44$, $T_2/T_1=4$, $T^{**}/T_1=2.25$. The reaction rate is the reciprocal of the mean first passage time. (a) Connecting region is extremely small: only one percent of the total region. $k_{AB}^1$ with $T^{**}$ and $k_{AB}^2$ with $T^{**}$ overlap and  approximate the exact results very well. (b) Connecting region is small but not extremely small: fifteen percent of the total region. $k_{AB}^1$ with $T^*$ and $k_{AB}^2$ with $T^*$ are more accurate. (c) Connecting region is large: forty percent of the total region. $k_{AB}^2$, no matter with $T^*$ or $T^{**}$, fails to fit the exact result, $k_{AB}^1$ with $T^*$ is a more accurate approximation. The Kramers' rate formula(the green line) doesnot approximate well in all cases.}
\label{validate_formula}
\end{figure*}

\begin{table}%The best place to locate the table environment is directly after its first reference in text
\caption{\label{table1}%
Comparison of different reaction-rate formulas in one dimension.}
\begin{ruledtabular}
\begin{tabular}{cc}
\textrm{Reaction-rate formula}&
\textrm{Connecting region}
\\
\colrule

$k_{AB}^1(T^{*})$  & not extremely narrow  \\
$k_{AB}^1(T^{**})$& extremely narrow  \\
$k_{AB}^2(T^{*})$  & narrow   \\
$k_{AB}^2(T^{**})$ & extremely narrow \\
\end{tabular}
\end{ruledtabular}
\end{table}

In one dimensional case, we assume that near $x_{A}$, $x_{B}$ and $x_{C}$ (Fig. \ref{barrier-crossing}),
\begin{eqnarray}
\begin{cases}
H(x)\approx H(x_{A})+\frac{1}{2}\lambda_{A}(x-x_{A})^{2},  \quad\quad x\approx x_{A};  \\
H(x)\approx H(x_{B})+\frac{1}{2}\lambda_{B}(x-x_{B})^{2},  \quad\quad x\approx x_{B};  \\
H(x)\approx H(x_{C})+\frac{1}{2}\lambda_{C}(x-x_{C})^{2},  \quad\quad x\approx x_{C},  \\
\end{cases}
\end{eqnarray}
in which $\lambda_{A},\lambda_{B}>0$ and $\lambda_C<0$.

When the thermal fluctuation is rather small, we have the following approximation (denoted as $k_{AB}^1$) for the transition rate from state $A$ to state $B$ (see Section IV of Supplementary Material \cite{supplementary2016material}):
\begin{equation}\label{rf1}
k_{AB}^1\approx\frac{\sqrt{-\lambda_A\lambda_C}}{2\pi\eta}\sqrt{\frac{T_{1}}{T^{*}}}e^{-\int_{x_A}^{x_C}\frac{H'(x)}{k_BT(x)}dx}.
\end{equation}

Now if the connecting region in Fig. \ref{barrier-crossing} is quite small, Eq.~(\ref{rf1}) can be simplified to the following approximation (denoted as $k_{AB}^2$)\cite{supplementary2016material}:
\begin{equation}\label{rf2}
k_{AB}^2\approx\frac{\sqrt{-\lambda_A\lambda_C}}{2\pi\eta}\sqrt{\frac{T_{1}}{T^{*}}}e^{-\frac{H(x_C)-H(x_A)}{k_{B}T_1}},
\end{equation}
in which $T^*=T(x_C)$.

Compared with Kramers' rate formula (denoted as $k_{kramers}$) \cite{mccann1999thermally}, we know
\begin{equation}\label{relation_Kramers}
k^2_{AB}=\sqrt{T_1/T^*}k_{kramers}.
\end{equation}

Furthermore, if the connecting region is extremely small, the temperature $T^*$ has little influence on the transition rate, so the $T^*$ in Eq.~(\ref{rf1}) and Eq.~(\ref{rf2}) should both be replaced by an alternative temperature $T^{**}$, which satisfies \cite{supplementary2016material}:
\begin{equation}\label{T^{**}}
\sqrt{T^{**}}=\frac{\sqrt{T_1}+\sqrt{T_2}}{2}.
\end{equation}

The conditions under which the revised rate formulas Eq.~(\ref{rf1}) and Eq.~(\ref{rf2}) approximate  the exact reaction rate are summarized in Table \ref{table1}, and numerical validation is shown in Fig. \ref{validate_formula}. The isothermal Kramers' rate formula approximates worse than the revised ones in all cases (Fig. \ref{validate_formula}).

Under isothermal condition, the revised reaction rates in Eq.~(\ref{rf1}) and Eq.~(\ref{rf2}) are equivalent, and also they are same with the established Kramer's rate formula. We will check whether these rate formulas are thermodynamic consistent or not using the high-dimensional versions below which are more general.

\subsubsection{High dimensional case}

\begin{figure}[hbt]
\centering
\subfigure[]{\includegraphics[width=7cm]{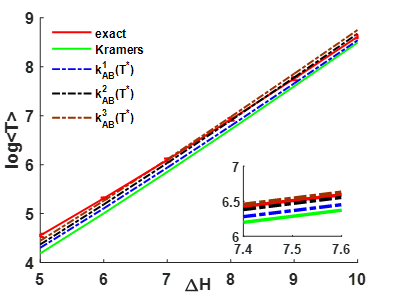}}
\caption{Validate the revised rate formulas under non-isothermal condition in the two dimensional case. The exact reaction rate is obtained via numerical simulations. The enthalpy potential function used is $H(x_1,x_2)=c*(\frac{1}{2}x_1^4-x_1^2)+x_2^2$. $T^*/T_1=1.44$, $T_2/T_1=1.96$, $\Delta H=0.5c$. The reaction rate is the reciprocal of the mean first passage time $<T>$. Connecting region is small. Both Eq.~\ref{rfm2}(minimum enthalpy) and Eq.~\ref{rfm3}(average enthalpy) are accurate approximations of the exact result when the connecting region is small. The Kramers' rate formula(the green line) doesnot approximate well.}
\label{validate_formula1}
\end{figure}

\begin{table*}
\caption{\label{table2}%
Comparison of different reaction-rate formulas in high dimension.}
\begin{ruledtabular}
\begin{tabular}{lcc}

\textrm{Reaction rate formula}&
\textrm{Connecting region}&
\textrm{Thermodynamically consistent}
\\
\colrule
$k_{AB}^1(T^{*})$ & not extremely narrow & yes \\

$k_{AB}^2(T^{*})$& narrow but not extremely narrow& {\bf no} \\

$k_{AB}^3(T^{*})$& narrow but not extremely narrow& yes\\

\end{tabular}
\end{ruledtabular}
\end{table*}

Without loss of generality, we denote the one-dimensional coordinate existing in the high-dimensional case as $x$ and the rest coordinates as $\mathbf{y}$. Note that $x\in R^1$ and $\mathbf{y}\in R^{n-1}$. $\{(x,\mathbf{y}):\mathbf{y}\in R^{n-1}\}$ is a set of parallel planes, perpendicular to the $x$ coordinate. We assume, for fixed $x$, $(x,0)$ is the minimum of the potential energy $H(x,y)$ on the plane $\{(x,\mathbf{y}):\mathbf{y}\in R^{n-1}\}$. We further assume, around reactant region $x=x_{A}$,  $H(x,\mathbf{y})\approx H(x_{A},\mathbf{0})+(x,\mathbf{y})^T\mathbf{A_n}(x_{A})(x,\mathbf{y})/2$, and around the transition state region $x=x_{C}$, $H(x,\mathbf{y})\approx H(x_{C},\mathbf{0})+(x,\mathbf{y})^T\mathbf{A_n}(x_{C})(x,\mathbf{y})/2$, where $\mathbf{A_n}(x)$ is the $n$-dimensional Hessian matrix at $(x,\mathbf{0})$. Further denote $\mathbf{A_{n-1}}(x_{C})$ as the $n-1$ dimensional Hessian matrix without the dimension of $x$. We have $\det \mathbf{A_n}(x_{C})=\lambda_{C}\det \mathbf{A_{n-1}}(x_{C})$, where $\lambda_{C}<0$ is the eigenvalue along the direction $x$.

When the thermal fluctuation is rather small, we can have the following approximation for the transition rate from state $A$ to state $B$ (see Section V of Supplementary Material \cite{supplementary2016material}):
{\small\begin{equation}\label{rfm1}
k_{AB}^1 \approx\frac{\sqrt{-\lambda_{C}}}{2\pi\eta}\sqrt{\frac{T_{1}}{T^{*}}}\left(\frac{\det \mathbf{A_{n}}(x_A)}{\det \mathbf{A_{n-1}}(x_C)}\right)^{\frac{1}{2}}e^{-\int_{x_A}^{x_C}\frac{H_{x}'(x,\mathbf{0})}{k_{B}T(x)}dx},
\end{equation}}
which is the high-dimensional version of Eq. (\ref{rf1}).

Then the entropy production after the completion of a counterclockwise cycle in Fig. \ref{four-state-model} is (see Section VI of Supplementary Material \cite{supplementary2016material})
{\small\begin{eqnarray}
&&k_B\ln\left(\frac{k_{12}k_{23}k_{34}k_{41}}{k_{14}k_{43}k_{32}k_{21}}\right)\\
&=&-\oint\frac{H_x'(x)}{T(x)}dx+\frac{\mu_{M_i}}{T_1}-\frac{\mu_{M_o}}{T_2},\nonumber
\end{eqnarray}}
which is exactly the general form of thermodynamic relation [Eq.~(\ref{e2})]. Thus Eq.~(\ref{rfm1}) is thermodynamic consistent.

If we assume that the connecting region around the transition state $x_C$ is quite narrow and let $T^*=T(x_C)$, Eq.~(\ref{rfm1}) can be simplified to
{\small\begin{equation}\label{rfm2}
k^2_{AB}\approx\frac{\sqrt{-\lambda_{C}}}{2\pi\eta}\sqrt{\frac{T_{1}}{T^{*}}}\left(\frac{\det \mathbf{A_{n}}(x_A)}{\det \mathbf{A_{n-1}}(x_C)}\right)^{\frac{1}{2}}e^{-\frac{H(x_C,0)-H(x_A,0)}{k_{B}T_1}},
\end{equation}}
which is the high-dimensional version of Eq. (\ref{rf2}) and satisfies Eq.~(\ref{relation_Kramers}).

However, even under local Gaussian approximation, the rate formula Eq.~(\ref{rfm2}) is not thermodynamic consistent with Eq.~(\ref{e1}) (see Section VI of Supplementary Material \cite{supplementary2016material}). Therefore, we need a new approximation to Eq.~(\ref{rfm1}).

Under the local Gaussian approximation, we derive another approximation of Eq.~(\ref{rfm1}) \cite{supplementary2016material}:
{\small\begin{equation}\label{rfm3}
k^3_{AB}\approx\frac{\sqrt{-\lambda_{C}}}{2\pi\eta e}\left(\frac{T^{*}}{T_{1}}\right)^{n-3/2}\left(\frac{\det \mathbf{A_{n}}(x_A)}{\det \mathbf{A_{n-1}}(x_C)}\right)^{\frac{1}{2}}e^{-\frac{H_C-H_A}{k_{B}T_1}}.
\end{equation}}
in which
{\small\begin{equation}
H_A\approx H(x_A,0)+nk_BT_1, \quad H_C\approx H(x_C,0)+(n-1)k_BT^*.
\end{equation}}
Notice that in calculating $H_C$ the reaction coordinate $x$ is left out, and $H_A$ involves one more coordinate than $H_C$ \cite{Zhou2010}.

Since the entropy of state $A$ and state $C$ under the local Gaussian approximation become \cite{Zhou2010}
{\small\begin{eqnarray}
&&S_C^*(T^*)\nonumber\\
&&=(n-1)k_{B}\ln\left(\frac{2\pi ek_BT^*\sqrt{m}}{h}\right)-\frac{1}{2}k_B\ln(\det \mathbf{A_{n-1}}(x_C))\nonumber\\
&&S_{A}(T_1)\nonumber\\
&&=nk_{B}\ln\left(\frac{2\pi ek_BT_1\sqrt{m}}{h}\right)-\frac{1}{2}k_B\ln(\det \mathbf{A_n}(x_A)),
\end{eqnarray}}

We rewrite the revised transition rate formula Eq.~(\ref{rfm3}) in terms of thermodynamic quantities
\begin{equation}\label{revised1}
k^3_{AB}=\kappa_{AB}\frac{k_{B}T_{1}^{3/2}}{h(T^*)^{1/2}}e^{-\frac{H_C-H_A}{k_{B}T_1}}e^{\frac{S_{C}^{\ddag}(T^*)-S_{A}(T_1)}{k_B}},
\end{equation}
in which $\kappa_{AB}=\frac{\sqrt{-\lambda_Cm}}{\eta}$ is the transmission coefficient and $h$ is the Planck constant. We can also consider the transition rate from state $B$ back to state $A$, then the prefactors $\kappa_{AB}=\kappa_{BA}=\frac{\sqrt{-\lambda_Cm}}{\eta}$.

Applying Eq.~(\ref{rfm3}), the entropy production along the counterclockwise cycle in Fig. \ref{four-state-model} becomes \cite{supplementary2016material}
{\small\begin{eqnarray}\label{gamma}
e_p&=&k_{B}\ln\left(\frac{k_{12}k_{23}k_{34}k_{41}}{k_{14}k_{43}k_{32}k_{21}}\right)\nonumber\\
&=&\frac{H_{14}^{*}-H_{23}^{*}+\mu_{M_i}}{T_1}+\frac{H_{32}^{*}-H_{41}^{*}-\mu_{M_o}}{T_2}.
\end{eqnarray}}
which is exactly the new thermodynamic relation[Eq.~(\ref{e1})]. Thus Eq.~(\ref{rf2}) is thermodynamic consistent.

The three approximations $k^1_{AB},k^2_{AB},k^3_{AB}$ are equivalent under isothermal condition, which is exactly the Kramers' rate formula. The first revised rate formula $k_{AB}^1$ (Eq.~\ref{rfm1}) holds even when the connecting region is large and is consistent with the thermodynamic relation Eq.~(\ref{e2}); the second revised rate formula $k_{AB}^2$ (Eq.~\ref{rfm2}) is a modified version of Kramers' rate formula (Eq.~\ref{relation_Kramers}), but is neither consistent with the thermodynamic relation Eq.~(\ref{e2}) nor Eq.~(\ref{e1}); the third revised rate formula $k_{AB}^3$ (Eq.~\ref{rfm3}) approximates better than the isothermal Kramers rate formula (Fig. \ref{validate_formula1}) and is consistent with the new thermodynamic relation(Eq.~\ref{e1}). The condition under which the three revised formulas holds and whether they are thermodynamically consistent are summarized in Table \ref{table2}.

When the connecting region is extremely narrow, $T^*$ should be replaced by $T^{**}$(Eq.~\ref{T^{**}}) in order to approximate better, which does not violate the thermodynamic relations either.

\section{A more complicated cotransporter model}

\begin{figure}[hbt]
\centering
\includegraphics[width=6cm]{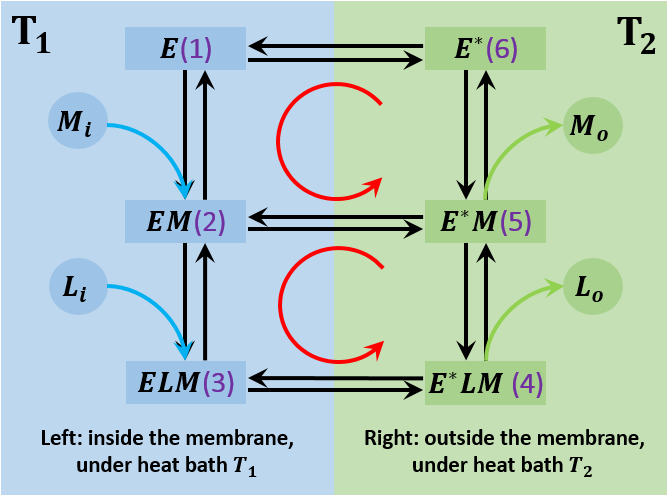}
\caption{Six-state non-isothermal model for the cotransporter across the membrane. To simplify the notations, we denote $E$ as state $1$, $EM$ as state $2$, $ELM$ as state $3$, $E^*LM$ as state $4$, $E^*M$ as state $5$ and $E^*$ as state $6$. $k_{ij}$ is the first-order or pseudo-first-order reaction constants from state $i$ to state $j$.}
\label{cotransporter}
\end{figure}

Fig.~\ref{cotransporter} shows a six-state model for the cotransporter across the membrane \cite{Hill1975,hill2012free}. We assume that a small molecule $M$ has a larger chemical potential inside the membrane than outside, i.e. $\mu_{M_i}> \mu_{M_o}$, and another small molecule $L$ has a larger chemical potential outside than inside, i.e. $\mu_{L_i}< \mu_{L_o}$. $M$ molecules would tend to move spontaneously from inside to outside whereas $L$ molecules would tend to move in the opposite direction.

There are three cycles in such a kinetic diagram (Fig.~\ref{cotransporter}). The thermodynamic analysis of the cycle $a$ (1-2-5-6-1) and cycle $b$ (2-3-4-5-2) are the same as the four-state model in Fig.~\ref{four-state-model}. For the cycle $a$ (1-2-5-6-1), the new thermodynamic relation similar to Eq.~(\ref{e1}) is
{\small$$k_{B}\ln \gamma_a=(H_{16}^{*}+\mu_{M_i}-H_{25}^{*})\left(\frac{1}{T_1}-\frac{1}{T_2}\right)+\frac{1}{T_2}(\mu_{M_i}-\mu_{M_o}),$$}
in which {\small$\gamma_a=\left(\frac{k_{12}k_{25}k_{56}k_{61}}{k_{21}k_{52}k_{65}k_{16}}\right)$}. The thermal driving force {\small$F^a_{thermo}=(H_{16}^{*}+\mu_{M_i}-H_{25}^{*})\left(\frac{1}{T_1}-\frac{1}{T_2}\right)$} and the chemical driving force {\small$F_{chem}^a=\frac{1}{T_2}(\mu_{M_i}-\mu_{M_o})$}.

Similarly for cycle $b$ (2-3-4-5-2), the new thermodynamic relation similar to Eq.~(\ref{e1}) is
{\small$$k_{B}\ln \gamma_b=(H_{25}^{*}+\mu_{L_i}-H_{34}^{*})\left(\frac{1}{T_1}-\frac{1}{T_2}\right)+\frac{1}{T_2}(\mu_{L_i}-\mu_{L_o}),$$}
in which {\small$\gamma_b=\left(\frac{k_{23}k_{34}k_{45}k_{52}}{k_{32}k_{43}k_{54}k_{25}}\right)$}. The thermal driving force {\small$F^b_{thermo}=(H_{25}^{*}+\mu_{L_i}-H_{34}^{*})\left(\frac{1}{T_1}-\frac{1}{T_2}\right)$} and the chemical driving force {\small$F_{chem}^b=\frac{1}{T_2}(\mu_{L_i}-\mu_{L_o})$}.

For the cycle $c$ (1-2-3-4-5-6-1) in which the transmembrane fluxes of molecules $M$ and $L$ are closely coupled, the new thermodynamic relation is just the summation of the previous two, i.e.
\begin{equation}
k_{B}\ln\gamma_c=F_{thermo}^c+F_{chem}^c,\nonumber
\end{equation}
in which $\gamma_c=\gamma_a\gamma_b$, $F_{thermo}^c=F_{thermo}^a+F_{thermo}^b$, and $F_{chem}^c=F_{chem}^a+F_{chem}^b$.

The thermodynamic relations for all three cycles are consistent with the revised reaction rate formulas Eq.~(\ref{rf2}) in the one-dimensional case and Eq.~(\ref{rfm3}) in the high-dimensional case. The more general thermodynamic relation similar to Eq.~(\ref{e2}) for each cycle can also be derived, which is consistent with the rate formulas Eq.~(\ref{rf1}) and Eq.~(\ref{rfm1}) (see Section VII of Supplementary Material \cite{supplementary2016material}).

Noticing the fact that {\small$\gamma_a=\frac{J^a_{+}}{J^a_{-}}$}, {\small$\gamma_b=\frac{J^b_{+}}{J^b_{-}}$} and {\small$\gamma_c=\frac{J^c_{+}}{J^c_{-}}$}, in which $J^a_{\pm}$, $J^b_{\pm}$ and $J^c_{\pm}$ are the averaged occurrences of forward and backward cycles $a$, $b$ and $c$ per unit time (called cycle fluxes) \cite{Hill1975,Jiang2004}, we know that in the absence of chemical driving force, i.e. $F_{chem}^a=F_{chem}^b=0$, or namely $\mu_{M_i}=\mu_{M_o}$ and $\mu_{L_i}=\mu_{L_o}$,
{\small
$$k_B\ln\frac{J^a_{+}}{J^a_{-}}=(H_{16}^{*}+\mu_{M_i}-H_{25}^{*})\left(\frac{1}{T_1}-\frac{1}{T_2}\right),$$
$$k_B\ln\frac{J^b_{+}}{J^b_{-}}=(H_{25}^{*}+\mu_{L_i}-H_{34}^{*})\left(\frac{1}{T_1}-\frac{1}{T_2}\right),$$
$$k_B\ln\frac{J^c_{+}}{J^c_{-}}=(H_{16}^{*}-H_{34}^{*}+\mu_{M_i}+\mu_{L_i})\left(\frac{1}{T_1}-\frac{1}{T_2}\right).$$
}
Therefore the net fluxes $J^a=J^a_{+}-J^a_{-}$, $J^b=J^b_{+}-J^b_{-}$ and $J^c=J^c_{+}-J^c_{-}$ satisfies
{\small\begin{eqnarray}
&&J^a\cdot (H_{16}^{*}+\mu_{M_i}-H_{25}^{*})\left(\frac{1}{T_1}-\frac{1}{T_2}\right)>0,\nonumber\\
&&J^b\cdot (H_{25}^{*}+\mu_{L_i}-H_{34}^{*})\left(\frac{1}{T_1}-\frac{1}{T_2}\right)>0,\nonumber\\
&&J^c\cdot (H_{16}^{*}-H_{34}^{*}+\mu_{L_i}+\mu_{M_i})\left(\frac{1}{T_1}-\frac{1}{T_2}\right)>0.
\end{eqnarray}}

Once $H_{16}^{*}+\mu_{M_i}-H_{25}^{*}>0$ and $H_{25}^{*}+\mu_{L_i}-H_{34}^{*}>0$, all the $J_a$, $J_b$ and $J_c$ have the same sign as {\small$\left(\frac{1}{T_1}-\frac{1}{T_2}\right)$}, indicating that all the net fluxes of these transported molecules go against the temperature gradient. It does not violate the Second Law.

The number of occurrences $\nu_{i}(t)$ of counterclockwise cycle $i$ ($i=a,b,c$) in Fig.~\ref{cotransporter} along the stochastic trajectory up to time $t$ satisfies the detailed fluctuation theorem \cite{Ge2012stochastic,Jiachen2016_a}

\begin{equation}
\frac{P(\nu_{i}(t)=k)}{P(\nu_{i}(t)=-k)}= \gamma_i^{k},
\end{equation}
for any integer $k$, $i=a,b,c$.

It is followed by \cite{Ge2012stochastic,Jiachen2016_a}
$$\langle e^{-\lambda \nu_{i}(t)}\rangle=\langle e^{-(\log\gamma_i-\lambda)\nu_i(t)}\rangle.$$
for any real $\lambda$.

Therefore, for any physical quantity associated with cycle $i$, $i=a,b,c$, denoted as $W(t)=C\cdot \nu_{i}(t)$ for any constant $C$, we can write
\begin{equation}
\frac{P(W(t)=w)}{P(W(t)=-w)}=\gamma_{i}^{w/C},
\end{equation}
for any $w$ in which $w/C$ is an integer, and
{\small\begin{equation}
\left\langle e^{-\lambda W(t)}\right\rangle=\left\langle e^{-\left(\frac{\log\gamma_{i}}{C}-\lambda\right)W(t)}\right\rangle.
\end{equation}}

\section{General master-equation model}

Consider a general master equation model for $N$ states $\{1,2,\cdots,N\}$,
\begin{equation}
\frac{dp_i(t)}{dt}=\sum_{j\neq i} \left(p_j(t)k_{ji}-p_i(t)k_{ij}\right),
\end{equation}
where $k_{ij}$ is the reaction rates from state $i$ to state $j$ and $p_i(t)$ is the probability of state $i$ at time $t$.

Suppose the temperature of state $i$ is $T_i$, and the enthalpy of the transition state along the reaction $i\rightarrow j$ is $H^*_{ij}$. For the cycle $c=(i_1\rightarrow i_2\rightarrow i_3\rightarrow\cdots\rightarrow i_n\rightarrow i_1)$, we assume that after the completion of such a cycle, there are $n_i$ molecules of chemical species $S_i$ at temperature $T_{S_i}$ being converted into $m_j$ molecules of chemical species $P_j$ at temperature $T_{P_j}$. Then the new thermodynamic relation for the cycle similar to Eq.~(\ref{e1}) should be
\begin{eqnarray}
&&k_{B}\ln \gamma^{(c)}=\sum_{k=1}^n \frac{H^*_{i_ki_{(k-1)}}-H^*_{i_ki_{(k+1)}}}{T_i}\nonumber\\
&&+\sum_i n_i\frac{\mu_{S_i}}{T_{S_i}}-\sum_j m_j\frac{\mu_{P_j}}{T_{P_j}},
\end{eqnarray}
in which the state $i_0$ is the same as the state $i_n$, and the state $i_{n+1}$ is the same as the state $i_1$, and $\gamma^{(c)}=\frac{k_{i_1i_2}k_{i_2i_3}\cdots k_{i_{n-1}i_n}k_{i_ni_1}}{k_{i_1i_n}k_{i_ni_{n-1}\cdots k_{i_3i_2}k_{i_2i_1}}}$ is the affinity of the cycle. It is consistent with the revised rate formulas Eq.~(\ref{rf2}) and Eq.~(\ref{rfm3}).

Also we can have a more general form of the new thermodynamic relation similar to Eq.~(\ref{e2})
\begin{eqnarray}
&&k_{B}\ln \gamma^{(c)}\\
&=&-\oint\frac{H_x'(x)}{T(x)}dx+\sum_i n_i\frac{\mu_{S_i}}{T_{S_i}}-\sum_j m_j\frac{\mu_{P_j}}{T_{P_j}},\nonumber
\end{eqnarray}
which is consistent with the revised rate formula Eq.~(\ref{rfm1}).

At steady state, the dynamics of the master equation can be decomposed into cycles. The ensemble averaged entropy production rate  \cite{Schnakenberg1976network,Jiang2004}
$$epr=\sum_{c:cycles} J^{c}\cdot k_{B}\ln \gamma^{(c)},$$
in which $J^{c}$ is the net occurrence of cycle $c$ in unit time and $J^{c}>0$ if and only if $\gamma^{(c)}>1$.

The number of occurrences $\nu_{c}(t)$ of forward cycle $c=(i_1\rightarrow i_2\rightarrow i_3\rightarrow\cdots\rightarrow i_n\rightarrow i_1)$ up to time $t$ satisfies the detailed fluctuation theorem \cite{Ge2012stochastic,Jiachen2016_a}
\begin{equation}
\frac{P(\nu_{c}(t)=k)}{P(\nu_{c}(t)=-k)}= \left(\gamma^{(c)}\right)^k,
\end{equation}
for any integer $k$. It is followed by \cite{Ge2012stochastic,Jiachen2016_a}
$$\langle e^{-\lambda \nu_{c}(t)}\rangle=\langle e^{-(\log\gamma^{(c)}-\lambda)\nu_c(t)}\rangle,$$
for any real $\lambda$.

Therefore, for any physical quantity associated with cycle $c$, denoted as $W(t)=C\cdot \nu_{c}(t)$ for any constant $C$, we can write
\begin{equation}
\frac{P(W(t)=w)}{P(W(t)=-w)}=\left(\gamma^{(c)}\right)^{w/C},
\end{equation}
for any $w$ in which $w/C$ is an integer, and
{\small\begin{equation}
\left\langle e^{-\lambda W(t)}\right\rangle=\left\langle e^{-\left(\frac{\log\gamma^{(c)}}{C}-\lambda\right)W(t)}\right\rangle.
\end{equation}}

\section{Conclusion and discussion}

In summary, we have introduced temperature difference as an additional driving force to the chemical potential difference in biochemical cycle kinetics. Under such a non-isothermal condition, the thermodynamic relation between the reaction rates and thermodynamic potentials should be modified. Our approach is based on the consistency of the macroscopic non-equilibrium thermodynamics along with the mesoscopic stochastic thermodynamics. Then we have derived several revised rate formulas for the single chemical reaction under the nonisothermal condition, most of which are consistent with the new thermodynamic relation and approximate the exact reaction rate better than the isothermal transition-state rate formula. The thermodynamic analysis here also suggests that the transmembrane flux of molecules in the cyclic transporter model can even go against the temperature gradient in absence of chemical driving force.

Several previous works have considered another non-isothermal setting, i.e. there are parallel reaction paths with different temperatures for each chemical reaction \cite{Dhar2008,Esposito2012,Qian2016}. This assumption is unrealistic if we describe the biochemical cycle kinetics inside a living cell by the master equation approach. The temperature in our theory is defined for chemical states rather than reaction transitions in the master-equation model, which forces us to consider more details of the reaction path, for instance the enthalpy and temperature of the transition state.

On the other hand, different parameter regions need different approximations. Even in the same parameter region, different approximations are still possible. Typically one can chose any of these approaches to be applied in the parameter region at which they are valid. But if we are also interested in the associated thermodynamics, then certain existing approximations might not be suitable, if it breaks down the thermodynamic laws. Reaction-rate formulas with simple expressions are always approximations to the exact values of reaction rates. Except how well the approximation is, its consistency with the thermodynamics is also an issue one should pay attention to, especially when we would like to apply these rate formulas studying the nonequilibrium thermodynamics in a concrete biochemical system.

Last but not the least, master equation is one kind of mathematical model for describing the stochastic biochemical dynamics, and if there is temperature gradient present, the vanishing of the entropy production rate of the master equation is not equivalent to the thermodynamic equilibrium: we need both the chemical and thermal driving forces vanish. The inconsistency between the vanishing of entropy production rate and thermodynamic equilibrium under the non-isothermal condition originates from the kinetic energy perspective of the temperature. In a description without kinetic energy, the thermal driving force can be hardly distinguished from other driving forces, when applying the mathematical framework of stochastic thermodynamics \cite{Celani2012,Spinney2012,Kwon2011,Ge2014}. It is why in the present work we need to carefully study the heat dissipation along the reaction coordinates of the involved chemical reactions.

\begin{acknowledgments}
The authors would like to thank Guangkuo Ai, Liang Luo and Das Biswajit for comments and helpful discussions. H. Ge
is supported by NSFC (No. 21373021 and 11622101), and the 863 program (No. 2015AA020406).
\end{acknowledgments}

\bibliography{nonisothermal_v15}

\end{document}